\documentclass[aps,pra,twocolumn,tighten,groupedaddress,floatfix]{revtex4-1}

\usepackage{graphicx}
\usepackage{dcolumn}
\usepackage{bm}
\usepackage{amsmath}
\usepackage{xcolor}
\usepackage{braket}

\usepackage{amsfonts}
\usepackage{blkarray}
\usepackage{amssymb}
\usepackage{multirow}
\usepackage{mathrsfs}

\DeclareMathOperator{\sgn}{sgn}

\begin{document}

\title{Anyon optics with time-of-flight two-particle \textcolor{black}{interference} of double-well-trapped 
interacting ultracold atoms}

\author{Constantine Yannouleas}
\email{Constantine.Yannouleas@physics.gatech.edu}
\author{Uzi Landman}
\email{Uzi.Landman@physics.gatech.edu}

\affiliation{School of Physics, Georgia Institute of Technology,
             Atlanta, Georgia 30332-0430}

\date{10 December 2018}

\begin{abstract}
The subject of bianyon interference with ultracold atoms is introduced through theoretical investigations
pertaining to the second-order momentum correlation maps of two anyons (built upon spinless and spin-$1/2$ 
bosonic, as well as spin-$1/2$ fermionic, ultracold atoms) trapped in a double-well optical trap. The 
two-particle system is modeled according to the recently proposed protocols for emulating an anyonic Hubbard 
Hamiltonian in ultracold-atom one-dimensional lattices. Because the second-order momentum correlations are 
mirrored in the time-of-flight second-order interference patterns in space, our findings provide impetus for 
time-of-flight experimental protocols for detecting anyonic statistics via interferometry measurements of massive 
particles that broaden the scope of the biphoton interferometry of quantum optics.   
\end{abstract}

\maketitle

\section{Introduction}
Emulations of condensed-matter many-body physics \cite{bloc08,zoll12} and of optical biphoton interferometry 
\cite{foel05,kauf14,kauf18,aspe15,isla15,bran17,bran18,yann19,bonn18} 
with ultracold atoms in optical traps and lattices,
as well as quantum simulations of many-body phenomena using nonlinear-optics platforms (e.g., coupled resonator 
arrays or waveguide lattices) \cite{hart06,gree06,ange07,fazi07,brom10,long12,lebu15,ange17} 
constitute complimentary branches of research that have 
witnessed explosive growth in the last two decades. A great promise of these emerging research branches rests with 
their potential for achieving actual simulations of exotic synthetic particles that have been theoretically proposed 
in many-body and elementary-particle physics, but have been problematic to realize within the experimental framework 
of traditional condensed-matter and high-energy subfields of physics.    

In this context, the properties and probable detection of synthetic particles, proposed initially in two
dimensions and referred to as anyons \cite{lein77,wilc82}, that obey nontrivial particle-exchange statistics
interpolating between the familiar bosonic and fermionic ones, continues to be an intensely active field of 
theoretical and experimental research across several disciplines of physics; see, e.g., in the context of quantum 
computing \cite{kita03,pach12}, 
current-current correlations of fractional-quantum-Hall anyons in high magnetic fields \cite{gefe12}, 
noninteracting ultracold anyonic atoms in harmonic traps \cite{dubc18}, and 
quasiholes in a fractional quantum Hall state of ultracold atoms \cite{umuc18}. 
We also note theoretical \cite{brom10,long12} and experimental \cite{lebu15} studies for 
simulating anyonic NOON states with photons in waveguide lattices.

Recently, going beyond the case of two-dimensional space, a propicious direction for the simulation of a new class 
of {\it massive\/} anyons opened when several experimental protocols (based on a fractional Jordan-Wigner 
transformation) were advanced 
\cite{keil11,gres15,ecka16}, showing that ultracold neutral atoms trapped in {\it onedimensional\/} optical lattices 
can offer an appropriate substrate for the implementation of anyonic statistics. In particular, an anyonic Hubbard 
model (related to spinless bosons) was formulated and, in analogy with condensed-matter themes, the 
influence of 1D anyonic statistics on ground-state phase transitions in extended optical lattices was explicitly 
studied in these \cite{keil11,gres15,ecka16} and subsequent publications \cite{pels15,fore16,zhan17}. 
\textcolor{black}{
Current interest in 1D anyonic Hubbard models remains expansive \cite{fesh17,gors18,zuo18,fore18}.  
}

Here, taking fully into account the interparticle interactions, we introduce the subject of 1D anyonic 
matter-wave two-particle interferometry with ultracold atoms and establish analogies with the quantum-optics 
biphoton \cite{mand99,shihbook,oubook} (two-photon coincidence) interferometry of massless and noninteracting 
photons. To this effect, in conforming with recent relevant experiments (which employ fermionic $^6$Li atoms 
\cite{joch15,joch18,prei18}), we present theoretical investigations of the second-order momentum 
correlation maps of three variants of a pair of anyons [built upon (i) spinless and (ii) spin-$1/2$ bosonic, as well
as (iii) spin-$1/2$ fermionic, ultracold atoms] trapped in an 
\textcolor{black}{
{\it isolated\/} optical-tweezer-created double well, serving as a twin-particle source for the subsequent 
time-of-flight (TOF) measurements. 
}

\textcolor{black}{
Going beyond the earlier spinless-bosons formalism \cite{keil11,gres15,ecka16}, this is achieved by our formulating 
anyonic Hubbard Hamiltonians that account for the spin-1/2 cases (ii) and (iii) above, in addition to the spinless
case (i). Because the second-order momentum correlations are mirrored in the TOF 
spectral maps in space \cite{altm04,yann19}, our findings provide a blueprint for TOF experimental 
protocols for probing anyonic statistics via second-order interferometry of massive particles that broaden the scope
of the biphoton \cite{mand99,shihbook,oubook} (referred to also as fourth-order) interferometry of quantum optics.
}

\textcolor{black}{
For experimental determinations of the above-noted second-order momentum correlations maps via 
TOF higher-order spectroscopy of trapped ultracold atoms (specifically of two fermionic $^6$Li atoms isolated
in a double-well optical-tweezer trap), see Refs.\ \cite{prei18,joch18}. In these experiments, after the tweezers' 
trapping is turned off, the short-range interactions have negligible effect and the flight of the two atoms is 
ballistic up to the far-field, where the coincidence measurement is performed utilizing a high-resolution camera.
To be noted is the fact that the {\it in situ\/} preparation of pre-expansion few-atom states is deterministic, 
i.e., with high certainty concerning the number $N$ of the few trapped atoms. Such deterministically prepared states 
correspond to pure eigenstates of the trapped few-atom system \cite{joch15}. 
}

\textcolor{black}{
To put the present work in the context of higher-order (second-order or higher) ultracold atom interferometry, 
we stress recent advances in the experimental processing of data and control and manipulation of ultracold atoms in 
colliding free-space beams or clouds (including free fall under the cloud's gravity) 
\cite{aspe07,hodg11,hodg13,kher13,aspe15,kher17}, as well as in  optical-lattice traps and isolated few-tweezer 
configurations (two or three atoms, {\it in situ\/} or TOF) \cite{foel05,kauf14,joch15,kauf18,prei18}. Such
developments have motivated a growing number of both experimental 
\cite{aspe07,hodg11,kher13,aspe15,kher17,foel05,kauf14,joch15,kauf18,joch18,prei18} and theoretical 
\cite{kher14,bran17,bran18,bonn18,yann19} studies concerning the analogies between second or higher-order 
quantum-optics interference \cite{mand99,shihbook,oubook} and matter-wave spectroscopy. Our study goes beyond the 
earlier established subfield of first-order atom interferometry \cite{carn91,prit09,bermbook,schl15}, 
akin to Young's one-photon which-way double-slit interference.   
}

One of the findings of our study is that the anyonic signature in the two-particle interferometry maps reflects the 
appearance of a generalized NOON state as a major component in the entangled wave function of the ultracold atoms 
trapped in the double well. This NOON-state component is of the form 
$(|2,0\rangle \pm e^{i\theta} |0,2\rangle)/\sqrt{2}$, where $\theta$ is the statistical angle determining the 
commutation (anticommutation) relations for the anyonic exchange (see below).      

\textcolor{black}{
The plan of the paper is as follows: 
In section II, we give a detailed discussion of the theoretical methodologies 
developed and used in this study. This includes a discussion of anyonic exchange, the fractional Jordan-Wigner 
transformation, and the density-dependent 1D anyonic Hubbard model Hamiltonian for the above-noted three cases, 
i.e., (i) spinless and (ii) spin-1/2 bosonic, as well as (iii) spin-1/2 fermionic ultracold atoms trapped in an 
isolated optical-tweezer-created double well. The analytic eigenvalues associated with the four solutions of the 
three Hubbard Hamiltonians are also displayed graphically (see Fig.\ 1).
In section III we give analytical results and graphical display (see Fig.\ 2) for second-order 
momentum correlation maps 
exhibiting signatures of anyonic statistics, that is dependence on the statistical angle, predicted from our model 
for the ground state and two of the excited states of a system comprising two interacting anyonic ultracold atoms 
trapped in a double well. The three above noted cases, (i)-(iii), are discussed under conditions of vanishing 
inter-particle interaction, as well as for strongly attractive and repulsive interactions. 
We briefly summarize in section IV.  Detailed analytical results are given in the Appendices. In Appendix A, 
we describe the solution for two bosonic-based spinless anyons, and in Appendix B the solution for two 
spin-1/2 anyons (whether bosonic- or fermionic-based) is given. The analytical results for second-order momentum 
correlation maps are derived in Appendix C, and in Appendix D we display (in Fig.\ 3) plots of the correlation maps 
for the excited state with energy $E_3$, complementing those shown in Fig.\ 2  (in section III), where the 
correlations maps for $E_1$, $E_2$, and $E_4$ where shown.  
}

\section{Theory preliminaries}

\subsection{Anyonic exchange}
For spin-$1/2$ (i.e., two-flavor) anyons, the annihilation and creation operators are denoted as $a_{j,\sigma}$ 
and $a_{j,\sigma}^\dagger$, where the index $j=1,2$ (or equivalently $j=L,R$) denotes the left-right well 
(corresponding Hubbard-model site). These operators obey anyonic commutation or anticommutation relations
\begin{align}
\begin{split}
a_{j,\sigma} a_{k,\sigma^\prime}^\dagger \mp & e^{-i \theta \sgn(j-k)} a_{k,\sigma^\prime}^\dagger a_{j,\sigma}
=\delta_{j,k}\delta_{\sigma,\sigma^\prime},\\
a_{j,\sigma} a_{k,\sigma^\prime}  \mp & e^{i \theta \sgn(j-k)} a_{k,\sigma^\prime} a_{j,\sigma} =0.
\end{split}
\end{align} 
The upper sign (commutation) applies for bosonic-based anyons; the lower sign (anticommutaion) for 
fermionic-based anyons. $\sgn(j-k)=1$ for $j>k$, $\sgn(j-k)=-1$ for $j<k$, and $\sgn(j-k)=0$ for $j=k$.
For bosonic-based {\it spinless\/} anyons, one drops the spin index $\sigma$.
On the same site, the two particles retain the usual bosonic or fermionic commutation relations.

\subsection{Case (i): Density-dependent Hubbard Hamiltonian for bosonic-based spinless anyons}

Adapting the many-site case of Refs.\ \cite{keil11,gres15,ecka16}, a two-site anyonic Hubbard Hamiltonian 
for bosonic-based spinless anyons is written as follows:
\begin{align}
H_{\rm spinless}=-J (a_L^\dagger a_R + a_R^\dagger a_L) + \frac{U}{2} \sum_{j=L,R} n_j(n_j-1), 
\label{hasl}
\end{align}
where $J$ is the tunneling parameter, $U$ is the on-site interaction parameter (repulsive or attractive), and 
$n_j=a_j^\dagger a_j$ is the number operator.

Using a fractional Jordan-Wigner transformation \cite{keil11}, 
\begin{align}
a_L=b_L\;\;{\rm and}\;\; a_R=b_R \exp(-i \theta n_L),
\end{align}
where $b_j$ describes a usual bosonic operator and $n_j=b_j^\dagger b_j=a_j^\dagger a_j$,
the anyonic Hamiltonian in Eq.\ (\ref{hasl}) is mapped onto a bosonic Hubbard Hamiltonian with 
occupation-dependent hopping from right to left, i.e.,
\begin{align}
H^B_{\rm spinless}=-J (b_L^\dagger b_R e^{-i\theta n_L}+ {\rm h.c.}) + \frac{U}{2} \sum_{j=L}^R n_j(n_j-1).
\label{hbsl}
\end{align}
For two particles, if the left (target) site in unoccupied, the tunneling parameter is simply $-J$. If it is
occupied by one boson, this parameter becomes $-Je^{-i\theta}$. 

\subsection{Case (ii): Density-dependent Hubbard Hamiltonian for bosonic-based spin-$1/2$ anyons}

In this case, we introduce a two-site anyonic Hubbard Hamiltonian for bosonic-based spin-$1/2$ anyons 
as follows:
\begin{align}
\begin{split}
& H_{\rm spin-1/2}^B = \\
& -J \sum_{\sigma} (a_{L,\sigma}^\dagger a_{R,\sigma}  + {\rm h.c.})
+  \frac{U}{2} \sum_{j=L,R} N_j(N_j-1),
\end{split}
\label{hasf}
\end{align}
where $N_j=\sum_\sigma a_{j,\sigma}^\dagger a_{j,\sigma}$, with $\sigma$ denoting the up ($\uparrow$) or down
($\downarrow$) spin; $N_j$ is the number operator at each site $j$ including the spin degree of freedom.

Using a modified fractional Jordan-Wigner transformation \cite{bati04}, 
\begin{align}
a_{L,\sigma}=b_{L,\sigma}\;\;{\rm and}\;\; a_{R,\sigma}=b_{R,\sigma} \exp(-i \theta N_L),
\end{align}
where $b_{j,\sigma}$ describes a usual spin-1/2 bosonic operator and $N_j=\sum_\sigma 
b_{j,\sigma}^\dagger b_{j,\sigma}= \sum_\sigma a_{j,\sigma}^\dagger a_{j,\sigma}$,
the anyonic Hamiltonian in Eq.\ (\ref{hasf}) is mapped onto a bosonic Hubbard Hamiltonian with 
occupation-dependent hopping from right to left, i.e.,
\begin{align}
\begin{split}
& H^B_{\rm spin-1/2}=\\
&-J \sum_\sigma (b_{L,\sigma}^\dagger b_{R,\sigma} e^{-i\theta N_L}+ {\rm h.c.}) + 
\frac{U}{2} \sum_{j=L,R} N_j(N_j-1).
\end{split}
\label{hbsf}
\end{align}
For two particles, if the left (target) site in unoccupied, the tunneling parameter is simply $-J$. If it is
occupied by one boson, this parameter becomes $-Je^{-i\theta}$. 

\subsection{Case (iii): Density-dependent Hubbard Hamiltonian for fermionic-based spin-$1/2$ anyons}

In this case, we introduce a two-site anyonic Hubbard Hamiltonian for fermionic-based spin-$1/2$ anyons 
as follows:
\begin{align}
H_{\rm spin-1/2}^F = -J \sum_{\sigma} (a_{L,\sigma}^{F\dagger} a^F_{R,\sigma}  + {\rm h.c.})
+  U \sum_{j=L,R} n^F_{j,\uparrow} n^F_{j,\downarrow},
\label{hafsf}
\end{align}
where $n^F_{j,\sigma}= a_{j,\sigma}^{F\dagger} a^F_{j,\sigma}$, with $\sigma$ denoting the up ($\uparrow$) 
or down ($\downarrow$) spin.

Using a modified fractional Jordan-Wigner transformation \cite{bati04}, 
\begin{align}
a^F_{L,\sigma}=f_{L,\sigma}\;\;{\rm and}\;\; a^F_{R,\sigma}=f_{R,\sigma} \exp(-i \theta N^F_L),
\end{align}
where $f_{j,\sigma}$ describes a usual spin-1/2 fermionic operator and $N^F_j=\sum_\sigma 
f_{j,\sigma}^\dagger f_{j,\sigma}= \sum_\sigma a_{j,\sigma}^{F\dagger} a^F_{j,\sigma}$,
the anyonic Hamiltonian in Eq.\ (\ref{hafsf}) is mapped onto a fermionic Hubbard Hamiltonian with 
occupation-dependent hopping from right to left, i.e.,
\begin{align}
\begin{split}
& H^F_{\rm spin-1/2}=\\
&-J \sum_\sigma (f_{L,\sigma}^\dagger f_{R,\sigma} e^{-i\theta N^F_L}+ {\rm h.c.}) + 
U \sum_{j=L,R} n^F_{j,\uparrow} n^F_{j,\downarrow}.
\end{split}
\label{hbfsf}
\end{align}
For two particles, if the left (target) site in unoccupied, the tunneling parameter is simply $-J$. If it is
occupied by one fermion, this parameter becomes $-Je^{-i\theta}$. 

\subsection{Matrix representation of Hamiltonians}
 
In order to solve the two-site two-particle problem specified by the Hubbard-type Hamiltonians in Eqs.\ 
(\ref{hbsl}), (\ref{hbsf}), and (\ref{hbfsf}), which have a density-dependent tunneling term, one needs to 
construct the corresponding matrix Hamiltonians. These matrices and the corresponding eigenenergies are presented 
below because for a finite number of particles they offer a better grasp of the role of the statistical angle 
$\theta$. The corresponding eigenvectors and other details of the derivation of the associated second-order
momentum correlations and interferometry maps are given in Appendices \ref{a1}-\ref{a3}. 
When $\theta=0$, these Hamiltonian
matrices reduce to the pure bosonic or fermionic two-trapped-particle interferometry problems; see Refs.\ 
\cite{bran17,bran18,yann19} for the pure fermionic interferometry case.

For spinless bosons, using the bosonic basis kets
\begin{align}
\ket{2,0},\;\; \ket{1,1},\;\;\ket{0,2},
\label{a1_1d}
\end{align}
where $\ket{n_L,n_R}$ (with $n_L+n_R=2$) corresponds to a permanent with $n_L$ ($n_R$) particles in the 
$L$ ($R$) site, one derives the following $3 \times 3$ matrix Hamiltonian associated with the anyonic Hubbard 
Hamiltonian in Eq.\ (\ref{hbsl})
\begin{align}
H=\left(
\begin{array}{cccc}
 U & -\sqrt{2} e^{-i\theta} J & 0 \\
 -\sqrt{2} e^{i\theta}J & 0 & -\sqrt{2} J  \\
 0 & -\sqrt{2} J & U  \\
\end{array}
\right).
\label{a11d}
\end{align}

The three eigenenergies of the matrix (\ref{a11d}) are given by 
\begin{align}
\begin{split}
E_1& = \frac{J}{2} ( {\cal U} - \sqrt{{\cal U}^2 +16}) \\
E_2& = J {\cal U} = U\\
E_3& = \frac{J}{2} ( {\cal U} + \sqrt{{\cal U}^2 +16}),
\end{split}
\label{enerd}
\end{align}
where ${\cal U}=U/J$; they are exact results and independent of the statistical angle $\theta$, unlike the 
mean-field energies \cite{keil11}. In contrast, the corresponding three normalized eigenvectors (see Appendix
\ref{a1}) do depend on the statistical angle $\theta$. As explicitly shown below, this dependence results in 
tunable anyonic signatures that can be detected with controlled experimental protocols.

\begin{figure}[t]
\includegraphics[width=8.0cm]{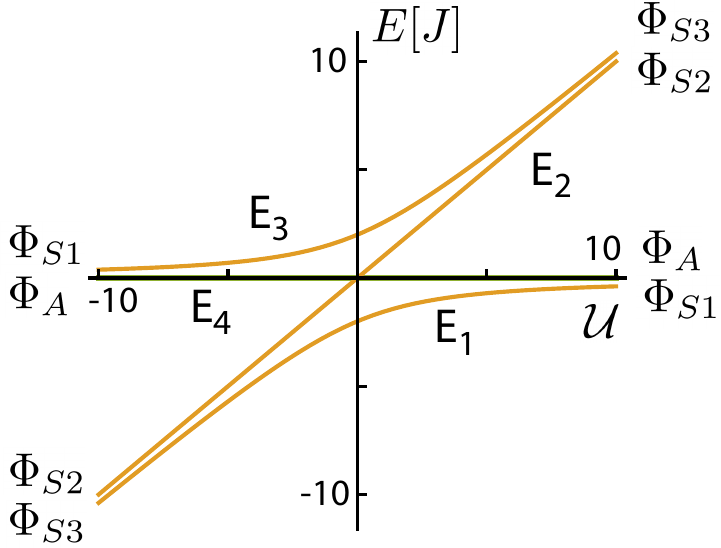}
\caption{Anyonic-Hubbard-dimer eigenenergies for all three cases of (i) spinless bosonic-based anyons, (ii) 
spin-1/2 bosonic-based anyons, and (iii) spin-1/2 fermionic-based anyons given by Eq.\ (13) plus $E_4=0$. 
The limiting $\Phi$ forms for the associated wave functions at ${\cal U} \rightarrow \pm \infty$ are also
denoted.
}
\label{fig1}
\end{figure}

For the two spin-1/2 cases (whether for two bosons or fermions), we seek solutions for states with $S_z=0$ 
\textcolor{black}{ (vanishing total-spin projection \cite{note}) }. 
In this case, the natural basis set is given by the four kets (note the choice of the ordering of these kets) 
\begin{align}
\ket{\uparrow\downarrow,0},\;\; \ket{\downarrow,\uparrow},\;\; \ket{\uparrow,\downarrow},
\;\;\ket{0,\uparrow\downarrow}.
\label{a1_4d}
\end{align}
In first quantization, these kets correspond to permanents for bosons and to determinants for fermions. Employing
this ket basis, one can derive the following $4 \times 4$ matrix Hamiltonians associated with the spin-$1/2$ Hubbard 
Hamiltonians in Eqs.\ (\ref{hbsf}) and (\ref{hbfsf}),
\begin{align}
H=\left(
\begin{array}{cccc}
 U & \mp e^{-i\theta} J & - e^{-i\theta} J & 0 \\
 \mp e^{i\theta} J & 0 & 0 & \mp J \\
 - e^{i\theta} J & 0 & 0 & -J \\
 0 & \mp J & -J & U
\end{array}
\right)
\label{a12d}
\end{align}
where the upper minus sign in $\mp$ applies for bosons and the bottom plus sign applies for fermions.

The four eigenenergies of the two matrices (\ref{a12d}) are given by the three quantities $E_i$, $i=1,\ldots,3$ 
in Eq.\ (\ref{enerd}) and an additional vanishing eigenenergy $E_4=0$; they are plotted in Fig.\ \ref{fig1} 
and they are are independent of the statistical 
angle $\theta$ and the $\mp$ alternation in sign. In contrast, as was also the case of the spinless bosons, the 
corresponding four normalized eigenvectors do depend on the statistical angle $\theta$; they are given in 
Appendix \ref{a2}.

\begin{figure*}[t]
\hspace{-0.6cm}\includegraphics[width=12cm]{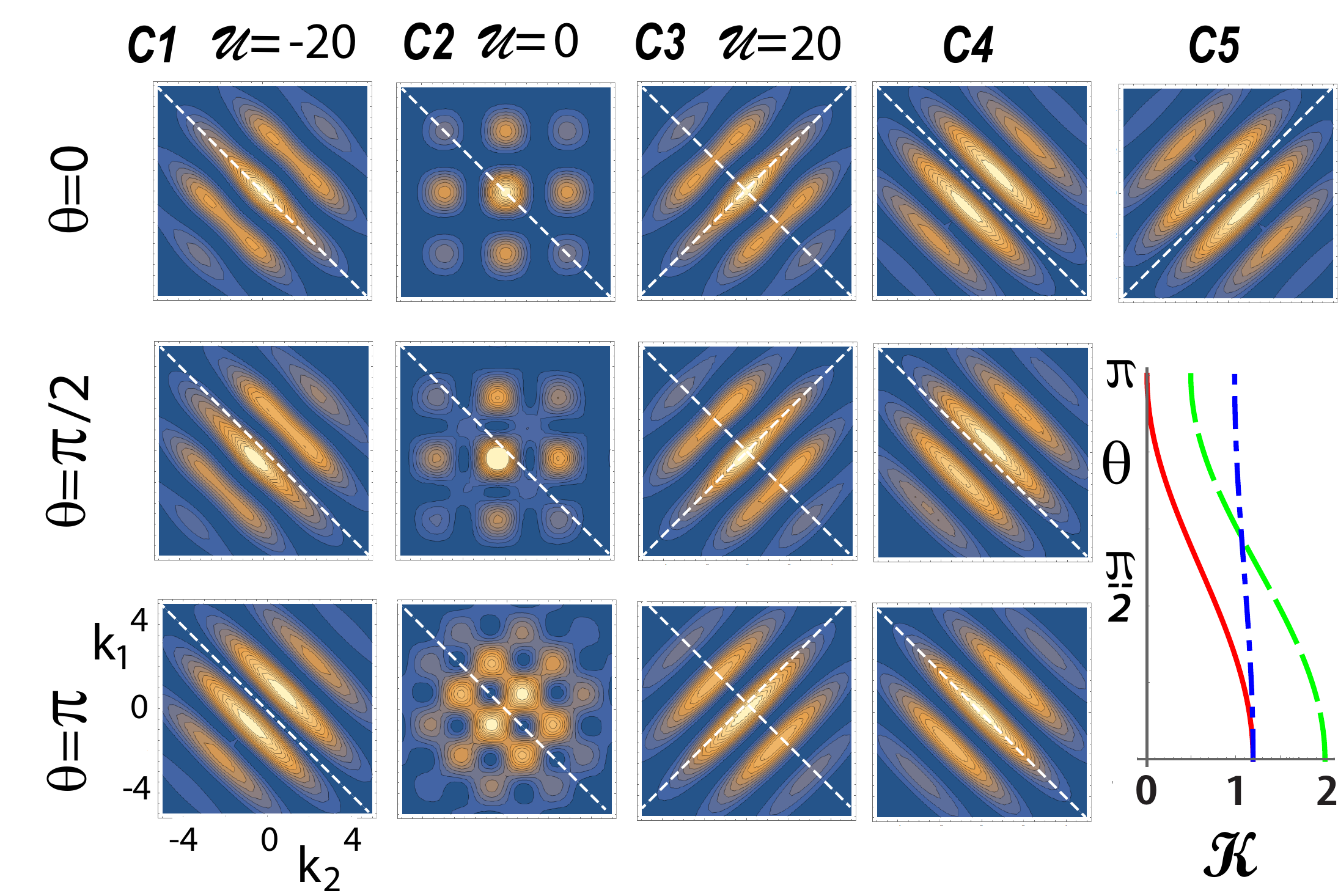}
\caption{Second-order momentum correlation maps exhibiting signatures of anyonic statistics (i.e., dependence on 
the statistical angle $\theta$) for two interacting anyonic ultracold atoms trapped in a double well. 
Columns $C1-C3$: case of the ground state (with energy $E_1$) [see Eq.\ (\ref{ggsd})], dependent on both the
interaction ${\cal U}$ and the statistical angle $\theta$. Column C1: strong attractive interparticle interaction 
${\cal U}=-20$. Column C2: vanishing interparticle interaction, ${\cal U}=0$. Column C3: strong repulsive 
interparticle interaction, ${\cal U}=20$.
Column $C4$: case of the excited state with energy $E_2$ [see Eq.\ (\ref{g2sd})], dependent on the statistical
angle $\theta$, but independent of the interaction  ${\cal U}$. Column $C5$, top frame: case of the excited state 
with energy $E_4=0$ [see Eq.\ (\ref{g4ad})], being independent from both  $\theta$ and ${\cal U}$; 
the wave function of this state is antisymmetric under the exchange of $k_1$ and $k_2$. Column $C5$, 
bottom frame: The functions ${\cal K}(\theta)=\pi {\cal G}_1^S(0,0,\theta)/(4s^2)$ that correspond to Figs.\ 
\ref{fig2}(C1) (red solid line), Figs.\ \ref{fig2}(C2) (green dashed line), and Figs.\ \ref{fig2}(C3) (blue 
dash-dotted line) for the ground state.
Top row: $\theta=0$ (pure bosons or fermions). Middle row: $\theta=\pi/2$ (intermediate anyons). Bottom row: 
$\theta=\pi$ (hard bosons or pseudofermions). The terms hard bosons and pseudofermions reflect the fact that the
onsite commutation (anticommutaion) relations do not change as a function of $\theta$, i.e., the onsite 
exclusion-principle behavior does not transmute from bosonic to fermionic and vice versa. The remaining parameters 
are: interwell distance, $2d=2$ $\mu$m and width of single-particle orbital, $s=0.2$ $\mu$m. 
\textcolor{black}{
$s$ governs the decay of the interference pattern away from the center of the map, while $1/d$ controls the
spacing between the fringes.
} 
$k_1$ and $k_2$ in units of $1/\mu$m. The dashed white lines are a guide to the eye. Blue represents the zero 
of the color scale. The white color corresponds to the maximum value of ${\cal G}(k_1,k_2,\theta)$.
(Blue is rendered into black in the printed version.)
}
\label{fig2}
\end{figure*}

\section{Results: Second-order momentum correlation maps}
\textcolor{black}{
The spatial far-field interference patterns map linearly onto the second-order momentum correlations characterizing 
the pure state of the atoms in the source (that is, in  the optical-tweezers-generated double-well confinement).
}
To generate the second-order momentum correlation maps ${\cal G}_i(k_1,k_2,\theta)$, $i=1,\ldots,4$, one needs to 
transit to the first-quantization formalism, which uses position- or momentum-dependent site-localized orbitals, 
$\psi_L$ and $\psi_R$. 
To this effect, each pure bosonic or fermionic particle in either of 
the two wells is represented by a displaced Gaussian function \cite{bran17,bran18,yann19}, which equivalently in 
momentum space is given by
\begin{equation}
\psi_{j}(k)=\frac{2^{1/4}\sqrt{s}}{\pi^{1/4}} e^{-k^2s^2} e^{id_j k},
\label{psikd}
\end{equation}
where again the index $j$ stands for $L$ (left) or $R$ (right); the separation between the two wells is
$2d=d_R-d_L$. 
\textcolor{black}{
The value of the single-particle spatial-extent parameter $s$, as well as the separation $2d$ between the wells are 
taken in the numerical illustrations (see Fig.\ \ref{fig2}) to have values (0.2 $\mu$m and 2 $\mu$m, respectively) 
similar to those used in experimental investigations of 1D trapped ultracold atoms \cite{prei18}.
}

The details of the derivation are given in Appendix \ref{a3}. Here we list the final analytical formulas for
the ${\cal G}_i(k_1,k_2,\theta)$'s, which are independent of the total spin (i.e., whether the state is spinless
or a spin singlet or a spin triplet state), and thus are the same for all three cases 
(i)-(iii). For the ground state, with energy $E_1$, one finds the following second-order momentum correlations
\begin{align}
\begin{split}
& {\cal G}_1^S(k_1,k_2,\theta) =  \frac{2s^2 e^{-2 s^2(k_1^2+k_2^2)}} { \pi \sqrt{{\cal U}^2+16} } \times \\
& \Big( {\cal R}({\cal U}) \cos^2[d(k_1-k_2)] + {\cal R}({-\cal U}) \cos^2[d(k_1+k_2)+\theta/2] + \\
& 8 \cos[d(k_1-k_2)] \cos[d(k_1+k_2)+\theta/2] \cos(\theta/2) \Big),
\end{split}
\label{ggsd}
\end{align}
where ${\cal R}({\cal U})=\sqrt{{\cal U}^2+16}+{\cal U}$.
\textcolor{black}{
The superscript $S$ here, and in Eqs.\ (\ref{g2sd}) and (\ref{g3sd}) below, denotes that the momentum part of the 
corresponding two-particle wave functions is symmetric under the exchange of the two momenta $k_1$ and $k_2$;
see Appendix \ref{a3}.
}

For the excited state with energy $E_2$, one finds the following second-order momentum correlations
\begin{align}
{\cal G}_2^S(k_1,k_2,\theta) = &  \frac{4s^2}{\pi} e^{-2s^2(k_1^2+k_2^2)} \sin^2[d(k_1+k_2)+\theta/2)].
\label{g2sd}
\end{align}

For the excited state with energy $E_3$, one finds the following second-order momentum correlations
\begin{align}
\begin{split}
& {\cal G}_3^S(k_1,k_2,\theta) =  \frac{2s^2 e^{-2 s^2(k_1^2+k_2^2)}} { \pi \sqrt{{\cal U}^2+16} } \times \\
& \Big( {\cal R}({-\cal U}) \cos^2[d(k_1-k_2)] + {\cal R}({\cal U}) \cos^2[d(k_1+k_2)+\theta/2] - \\
& 8 \cos[d(k_1-k_2)] \cos[d(k_1+k_2)+\theta/2] \cos(\theta/2) \Big),
\end{split}
\label{g3sd}
\end{align}

Finally, for the excited state with energy $E_4$ [only for the two spin-1/2 cases (ii) and (iii)], one finds the 
following second-order momentum correlations
\begin{align}
\begin{split}
{\cal G}_4^A(k_1,k_2,\theta) = \frac{4s^2}{\pi} e^{-2s^2(k_1^2+k_2^2)} \sin^2[d(k_1-k_2)].
\label{g4ad}
\end{split}
\end{align}
\textcolor{black}{
The superscript $A$ here denotes that the momentum part of the corresponding two-particle wave function is
antisymmetric under the exchange of the two momenta $k_1$ and $k_2$; see Appendix \ref{a3}.
}

The ${\cal G}_i(k_1,k_2,\theta)$ expressions above exhibit the following properties:
\textcolor{black}{
(1) The first three ${\cal G}_i$'s ($i=1,2,3$) are associated with two-particle eigenstates whose momentum parts are
symmetric under the exchange of the two momenta $k_1$ and $k_2$. 
Consequently, the underlying nodal structure does not allow a zero valley along the main diagonal.} 
These three cases depend on the statistical angle $\theta$. 
Thus their time-of-flight measurement will provide a signature for anyonic
statistics. (2) The statistical angle $\theta$ appears only in conjunction with cosine or sine terms containing the 
sum $k_1+k_2$ in their arguments. Cosine or sine terms containing only the difference $k_1-k_2$ of the two momenta 
are independent of $\theta$. This is a reflection of the fact that the vector solutions of the anyonic matrix 
Hamiltonians [see Eqs.\ (A4) and (B3)] contain the phase $e^{i\theta}$ only in the NOON-state 
component \cite{brom10,long12,lebu15} (of the form $(|2,0\rangle \pm e^{i\theta} |0,2\rangle)/\sqrt{2}$ or 
$\ket{\uparrow\downarrow,0} \pm e^{i\theta} \ket{0,\uparrow\downarrow}$, see Appendices \ref{a1}-\ref{a2}), 
and not in the Einstein-Podolski-Rosen-state component \cite{shih03} (of the form $|1,1 \rangle$ or 
$\ket{\downarrow,\uparrow} \pm \ket{\uparrow,\downarrow}$). 
\textcolor{black}{
(4) Only the fourth one ($i=4$, corresponding to the 
constant energy $E_4=0$) is associated with a two-particle eigenstate whose momentum part is antisymmetric under the
exchange of $k_1$ and $k_2$; 
consequently, the undelying nodal structure enforces a zero valley along the main diagonal.} 
This state, which corresponds to two {\it indistinguishable\/} fermions (e.g., two $^6$Li atoms in a 
triplet excited state) or bosons, is devoid of anyonic statistics.  

Fig.\ \ref{fig2} displays three cases (corresponding to the ground state and the two excited states with energies
$E_2$ and $E_4$) of second-order momentum correlation maps that illustrate the above properties. 
Keeping with property (2)
above, the variation of the interference patterns as a function of $\theta$ are more intense the larger the 
${\cal U}$-dependent contribution of the $k_1+k_2$ terms in the total ${\cal G}$ (the $k_1+k_2$ contributions 
produce interference fringes parallel to the antidiagonal). We note the alternation from a ridge to a valley along
the antidiagonal in Fig.\ \ref{fig2}(C1) (ground state at attractive ${\cal U}=-20$) and vice versa in Fig.\ 
\ref{fig2}(C4) ($E_2$ state independent of ${\cal U}$). For the ground state in the absence of interactions [Fig.\
\ref{fig2}(C2)], visible modifications (as a function of $\theta$) of a plaid-type theme persist in the interference
patterns. For the case when the $k_1+k_2$ terms have a small (or vanishing) contribution, the variations of the maps
are minimal [see Fig.\ \ref{fig2}(C3)] [or are absent, see Fig.\ \ref{fig2}(C5), top frame]; in this case, the 
dominance of the $\theta$-independent $k_1-k_2$ contributing terms is reflected in fringes parallel to the main 
diagonal. The bottom frame in the C5 column offers a complementary view of the $\theta$ dependence by plotting the 
curves ${\cal K}(\theta)=\pi {\cal G}_1^S(k_1=0,k_2=0,\theta)/(4s^2)$ that correspond to Figs.\ \ref{fig2}(C1), 
Figs.\ \ref{fig2}(C2), and Figs.\ \ref{fig2}(C3) for the ground state.      

For completeness, the case of the excited state with energy $E_3$ is presented in Appendix \ref{a4}; see 
Fig.\ \ref{fig3}.

\section{Summary}

In summary, the paper introduced the subject of matter-wave interferometry of massive and interacting anyons that 
can be realized with trapped 1D ultracold atoms in optical lattices. Furthermore, it analyzed the pertinent
signatures in the framework of time-of-flight experiments, and it established analogies with 
the interferometry of massless and noninteracting photonic anyons in waveguide lattices 
\cite{brom10,long12,lebu15}.
In particular, for two ultracold-atom anyons in a double-well confinement, this analogy is reflected in the fact 
that the NOON-state component of the massive bianyon is also of the form 
$(|2,0\rangle \pm e^{i\theta} |0,2\rangle)/\sqrt{2}$, where $\theta$ is the statistical angle determining the 
commutation (anticommutation) relations for the anyonic exchange.

\acknowledgments

This work has been supported by a grant from the Air Force Office of Scientic Research (AFOSR, USA) under Award
No. FA9550-15-1-0519. Calculations were carried out at the GATECH Center for Computational Materials Science.

\appendix

\section{Solution for two bosonic-based spinless anyons} 
\label{a1}

Using the bosonic basis kets
\begin{align}
\ket{2,0},\;\; \ket{1,1},\;\;\ket{0,2},
\label{a1_1}
\end{align}
where $\ket{n_L,n_R}$ (with $n_L+n_R=2$) corresponds to a permanent with $n_L$ ($n_R$) particles in the 
$L$ ($R$) site, one derives the following matrix Hamiltonian associated with the anyonic Hubbard 
Hamiltonian in Eq.\ (\ref{hbsl})
\begin{align}
H=\left(
\begin{array}{cccc}
 U & -\sqrt{2} e^{-i\theta} J & 0 \\
 -\sqrt{2} e^{i\theta}J & 0 & -\sqrt{2} J  \\
 0 & -\sqrt{2} J & U  \\
\end{array}
\right).
\label{a11}
\end{align}

The three eigenenergies of the matrix (\ref{a11}) are given by 
\begin{align}
\begin{split}
E_1& = \frac{J}{2} ( {\cal U} - \sqrt{{\cal U}^2 +16}) \\
E_2& = J {\cal U} = U\\
E_3& = \frac{J}{2} ( {\cal U} + \sqrt{{\cal U}^2 +16}),
\end{split}
\label{ener}
\end{align}
where ${\cal U}=U/J$. These eigenenergies are plotted in Fig.\ \ref{fig1}.

The corresponding three normalized eigenvectors are
\begin{align}
\begin{split}
{\cal V}_1&=\{{\cal B}({\cal U}) e^{-i\theta} /\sqrt{2},\; 
{\cal A}({\cal U}),\; {\cal B}({\cal U})/\sqrt{2} \}^T\\
{\cal V}_2&=\{e^{-i\theta}/\sqrt{2},\; 0,\; -1/\sqrt{2} \}^T\\
{\cal V}_3&=\{{\cal E}({\cal U}) e^{-i\theta} /\sqrt{2},\; 
{\cal D}({\cal U}),\; {\cal E}({\cal U})/\sqrt{2} \}^T,
\end{split}
\label{a1_2}
\end{align}
where the coefficients ${\cal A}$, ${\cal B}$, ${\cal D}$, and ${\cal E}$ are given by
\begin{align}
\begin{split}
{\cal A}({\cal U}) & =
\frac{{\cal U}+\sqrt{{\cal U}^2+16}}{\sqrt{2}\sqrt{ {\cal U}^2 +{\cal U} \sqrt{{\cal U}^2+16}+16}}, \\
{\cal B}({\cal U}) & =
\frac{4}{\sqrt{2}\sqrt{ {\cal U}^2 +{\cal U} \sqrt{{\cal U}^2+16}+16}}, \\
{\cal D}({\cal U}) & = -{\cal A}(-{\cal U}), \\
{\cal E}({\cal U}) & = {\cal B}(-{\cal U}).
\end{split}
\label{abde}
\end{align}

\section{Solution for two spin-$1/2$ anyons\/} 
\label{a2}

We seek solutions for states with $S_z=0$ (vanishing total spin
projection). In this case, the natural basis set is given by the four kets (note the choice of the ordering of
these kets) 
\begin{align}
\ket{\uparrow\downarrow,0},\;\; \ket{\downarrow,\uparrow},\;\; \ket{\uparrow,\downarrow},
\;\;\ket{0,\uparrow\downarrow}.
\label{a1_4}
\end{align}
In first quantization, these kets correspond to permanents for bosons and to determinants for fermions. Employing
this basis, one can derive the following $4 \times 4$ matrix Hamiltonians associated with the spin-$1/2$ Hubbard 
Hamiltonians in Eqs.\ (\ref{hbsf}) and (\ref{hbfsf}),
\begin{align}
H=\left(
\begin{array}{cccc}
 U & \mp e^{-i\theta} J & - e^{-i\theta} J & 0 \\
 \mp e^{i\theta} J & 0 & 0 & \mp J \\
 - e^{i\theta} J & 0 & 0 & -J \\
 0 & \mp J & -J & U
\end{array}
\right)
\label{a12}
\end{align}
where the upper minus sign in $\mp$ applies for bosons and the bottom plus sign applies for fermions.

The four eigenenergies of the matrices (\ref{a12}) are given by the quantities $E_i$, $i=1,\ldots,3$ in Eq.\ 
(\ref{ener}) and $E_4=0$; they are independent of the $\mp$ alternation in sign. The corresponding four 
normalized eigenvectors are
\begin{align}
\begin{split}
{\cal V}_1&=\{{\cal B}({\cal U}) e^{-i\theta} /\sqrt{2},\; 
\pm {\cal A}({\cal U})/\sqrt{2},\; {\cal A}({\cal U})/\sqrt{2},\;
{\cal B}({\cal U})/\sqrt{2} \}^T\\
{\cal V}_2&=\{e^{-i\theta} /\sqrt{2},\; 0,\; 0,\; -1/\sqrt{2} \}^T\\
{\cal V}_3&=\{{\cal E}({\cal U}) e^{-i\theta} /\sqrt{2},\; 
\pm {\cal D}({\cal U})/\sqrt{2},\; {\cal D}({\cal U})/\sqrt{2},\;
{\cal E}({\cal U})/\sqrt{2} \}^T\\
{\cal V}_4&=\{0,\; 1/\sqrt{2},\; \mp 1/\sqrt{2},\; 0 \}^T,
\end{split}
\label{a1_5}
\end{align}
where the upper sign (in $\pm$ or $\mp$) applies for bosons and the bottom sign applies for fermions.

\section{Second-order momentum correlation maps}
\label{a3}

To generate the second-order momentum correlation maps, one needs to transit from the ket notation to the wave 
function notation by employing the single-particle momentum-dependent site-localized orbitals $\psi_L(k)$ and 
$\psi_R(k)$ given in Eq.\ (\ref{psikd}). Indeed, in the first representation, the kets correspond to 
permanents for bosons or to determinants for fermions made of the $\psi_L(k)$ and $\psi_R(k)$ orbitals. 

One finds the following correspondence for spinless anyons
\begin{align}
\begin{split}
&\ket{1,1} \rightarrow \Phi_{S1}(k_1,k_2)\\
&e^{-i\theta} \ket{2,0}-\ket{0,2} \rightarrow \sqrt{2} \Phi_{S2}(k_1,k_2,\theta)\\
& e^{-i\theta} \ket{2,0}+\ket{0,2} \rightarrow \sqrt{2} \Phi_{S3}(k_1,k_2,\theta),
\end{split}
\label{a1_3}
\end{align}
and 
\begin{align}
\begin{split}
&\ket{\uparrow,\downarrow} \pm \ket{\downarrow,\uparrow} \rightarrow \sqrt{2} \Phi_{S1}(k_1,k_2) {\cal X}_1 \\
&e^{-i\theta} \ket{\uparrow\downarrow,0}-
\ket{0,\uparrow\downarrow} \rightarrow \sqrt{2} \Phi_{S2}(k_1,k_2,\theta) {\cal X}_2 \\
& e^{-i\theta}\ket{\uparrow\downarrow,0}+
\ket{0,\uparrow\downarrow} \rightarrow \sqrt{2} \Phi_{S3}(k_1,k_2,\theta) {\cal X}_3 \\
&\ket{\uparrow,\downarrow} \mp \ket{\downarrow,\uparrow} \rightarrow \sqrt{2} \Phi_{A}(k_1,k_2) {\cal X}_4.
\end{split}
\label{a1_6}
\end{align}
for spin-$1/2$ anyons, where the upper sign applies to bosonic-based anyons and the bottom sign applies to 
fermionic-based ones. ${\cal X}_i=\chi(1,0)$ for $i=1,2,3$ and ${\cal X}_4=\chi(0,0)$ for bosons and
${\cal X}_i=\chi(0,0)$, $i=1,2,3$ and ${\cal X}_4=\chi(1,0)$ for fermions; $\chi(0,0)$ and $\chi(1,0)$ are
the singlet and triplet spin eigenfunctions, respectively. The $\Phi$ functions are as follows:
\begin{widetext}
\begin{align}
\begin{split}
& \Phi_{S1}(k_1,k_2) = \big( \psi_L(k_1)\psi_R(k_2)+\psi_R(k_1)\psi_L(k_2) \big)/\sqrt{2} =
\frac{2s}{\sqrt{\pi}} e^{-s^2(k_1^2+k_2^2)} \cos[d(k_1-k_2)], \\
& \Phi_{S2}(k_1,k_2,\theta) = \big( e^{-i\theta} \psi_L(k_1)\psi_L(k_2)-\psi_R(k_1)\psi_R(k_2) \big)/\sqrt{2} = 
-i \frac{2s}{\sqrt{\pi}} e^{-s^2(k_1^2+k_2^2)} e^{-i\theta/2} \sin[d(k_1+k_2)+\theta/2)], \\
& \Phi_{S3}(k_1,k_2,\theta) = \big( e^{-i\theta} \psi_L(k_1)\psi_L(k_2)+\psi_R(k_1)\psi_R(k_2) \big)/\sqrt{2} =
\frac{2s}{\sqrt{\pi}} e^{-s^2(k_1^2+k_2^2)} e^{-i\theta/2} \cos[d(k_1+k_2)+\theta/2)], \\
& \Phi_{A}(k_1,k_2) = \big( \psi_L(k_1)\psi_R(k_2)-\psi_R(k_1)\psi_L(k_2) \big)/\sqrt{2} =
-i \frac{2s}{\sqrt{\pi}} e^{-s^2(k_1^2+k_2^2)} \sin[d(k_1-k_2)].
\end{split}
\label{phis}
\end{align}

For the ground state, with energy $E_1$, one finds the following second-order momentum correlations
\begin{align}
\begin{split}
& {\cal G}_1^S(k_1,k_2,\theta) = 
| {\cal A}({\cal U}) \Phi_{S1}(k_1,k_2) + {\cal B}({\cal U}) \Phi_{S3}(k_1,k_2,\theta) |^2 = \\
& \frac{4s^2}{\pi} e^{-2 s^2(k_1^2+k_2^2)} \Big( {\cal A}({\cal U})^2 \cos^2[d(k_1-k_2)] +
{\cal B}({\cal U})^2 \cos^2[d(k_1+k_2)+\theta/2] + \\
& 2 {\cal A}({\cal U}) {\cal B}({\cal U}) \cos[d(k_1-k_2)] \cos[d(k_1+k_2)+\theta/2] \cos(\theta/2) \Big).
\end{split}
\label{ggs}
\end{align}

For the excited state with energy $E_2$, one finds the following second-order momentum correlations
\begin{align}
\begin{split}
 {\cal G}_2^S(k_1,k_2,\theta) = |\Phi_{S2}(k_1,k_2,\theta)|^2 = 
  \frac{4s^2}{\pi} e^{-2s^2(k_1^2+k_2^2)} \sin^2[d(k_1+k_2)+\theta/2)].
\end{split}
\label{g2s}
\end{align}

For the excited state with energy $E_3$, one finds the following second-order momentum correlations
\begin{align}
\begin{split}
& {\cal G}_3^S(k_1,k_2,\theta) =
| -{\cal A}(-{\cal U}) \Phi_{S1}(k_1,k_2) + {\cal B}(-{\cal U}) \Phi_{S3}(k_1,k_2,\theta) |^2 = \\
& \frac{4s^2}{\pi} e^{-2 s^2(k_1^2+k_2^2)} \Big( {\cal A}(-{\cal U})^2 \cos^2[d(k_1-k_2)] +
{\cal B}(-{\cal U})^2 \cos^2[d(k_1+k_2)+\theta/2] \\
& -2 {\cal A}(-{\cal U}) {\cal B}(-{\cal U}) \cos[d(k_1-k_2)] \cos[d(k_1+k_2)+\theta/2] \cos(\theta/2) \Big).
\end{split}
\label{g3s}
\end{align}

Finally, for the excited state with energy $E_4=0$, one finds the following second-order momentum correlations
\begin{align}
\begin{split}
 {\cal G}^A_4(k_1,k_2,\theta) = |\Phi_{A}(k_1,k_2)|^2 = 
 \frac{4s^2}{\pi} e^{-2s^2(k_1^2+k_2^2)} \sin^2[d(k_1-k_2)].
\end{split}
\label{g4a}
\end{align}
~~~~~~\\
~~~~~~\\
\end{widetext}

\begin{figure}[t]
\hspace{-0.6cm}\includegraphics[width=7.6cm]{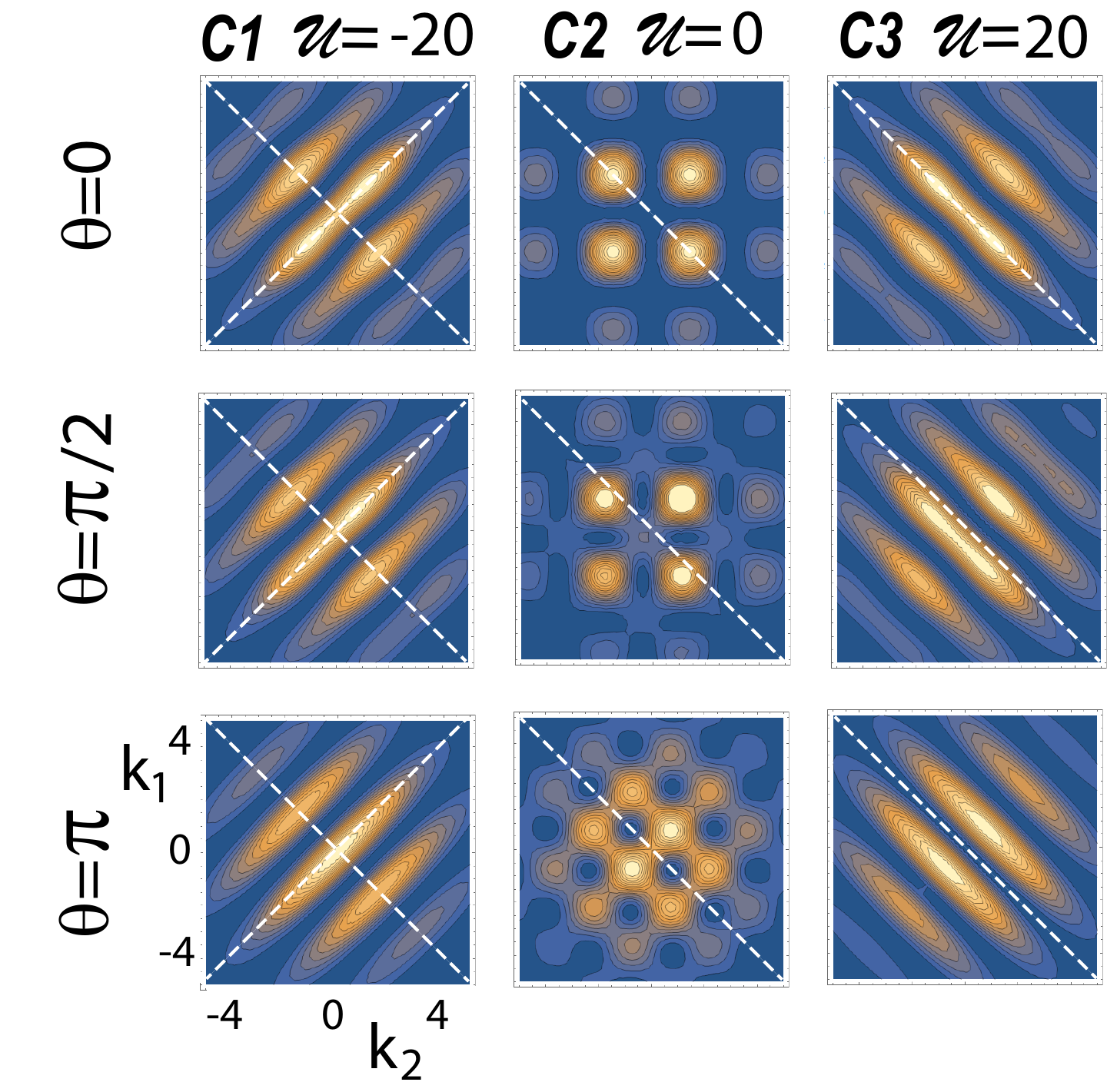}
\caption{Second-order momentum correlations of the excited state with energy $E_3$ of two interacting anyonic
ultracold atoms trapped in a double well [see Eq.\ (\ref{g3sd})], demonstrating dependence on the
statistical angle $\theta$. Top row: $\theta=0$ (pure bosons or fermions). Middle row: $\theta=\pi/2$ (intermediate
anyons). Bottom row: $\theta=\pi$ (hard bosons or pseudofermions). Column C1: attractive interparticle interaction
${\cal U}=-20$. Column C2: vanishing interparticle interaction, ${\cal U}=0$. Column C3: repulsive interparticle
interaction, ${\cal U}=20$.  The remaining parameters are: interwell distance, $2d=2$ $\mu$m and width of
single-particle orbital, $s=0.2$ $\mu$m. $k_1$ and $k_2$ in units of $1/\mu$m. The dashed white lines are a guide
to the eye. Blue represents the zero of the color scale. The white color corresponds to the maximum value of
${\cal G}_3^{S}(k_1,k_2,\theta)$. (Blue is rendered into black in the printed version.)
}
\label{fig3}
\end{figure}

With regard to the derivation of the expressions in Eqs.\ (\ref{ggs})$-$(\ref{g4a}), we note that,
generally, the second-order (two-particle) space density $\rho(x_1,x_1^\prime,x_2,x_2^\prime)$ 
for an $N$-particle system is defined as an integral over the product of the many-body wave function 
$\Psi(x_1, x_2, \ldots,x_N)$ and its complex conjugate $\Psi^*(x_1^\prime, x_2^\prime, \ldots,x_N)$, 
taken over the coordinates $x_3,\ldots,x_N$ of $N-2$ particles. To obtain the second-order space correlation 
function, ${\cal G}(x_1,x_2)$, one sets $x_1^\prime=x_1$ and $x_2^\prime=x_2$. The second-order momentum 
correlation function ${\cal G}(k_1,k_2)$ is obtained via a Fourier transform (from real space to momentum space) 
of the two-particle space density  $\rho(x_1,x_1^\prime,x_2,x_2^\prime)$ \cite{bran17,bran18}. In the case
of $N=2$, the above general definition reduces to a simple expression for the two-particle correlation 
functions, as the modulus square of the two-particle wave function itself; this applies in both cases whether 
the two-particle wave function is written in space or in momentum coordinates. 
This simpler second approach was followed here for deriving above the second-order momentum correlations for two
anyons.\\
~~~~\\
\section{Plots of correlation maps for the excited state with energy $E_3$}
\label{a4}

Fig.\ \ref{fig3} displays the second-order correlation maps for the excited state with energy $E_3$. It complements
Fig.\ \ref{fig2} where the corresponding maps for the three eigenstates with energies
$E_1$, $E_2$, and $E_4=0$ were displayed. For a description of these states as a function of the interparticle 
on-site interaction, ${\cal U}$,  see Fig.\ \ref{fig1}.


\begin{thebibliography}{99}
\bibitem{bloc08}
I. Bloch, J. Dalibard, and W. Zwerger,
Many-body physics with ultracold gases,
Rev. Mod. Phys. {\bf 80}, 885 (2008).
\bibitem{zoll12}
J.I. Cirac and P. Zoller,
Goals and opportunities in quantum simulation,
Nature Phys. {\bf 8}, 264–266 (2012).
\bibitem{foel05}
S. F\"{o}lling, F. Gerbier, A. Widera, O. Mandel, T. Gericke, and I. Bloch,
Spatial quantum noise interferometry in expanding ultracold atom clouds,
Nature {\bf 434}, 491 (2005).
\bibitem{kauf14}
A.M. Kaufman, B.J. Lester, C.M. Reynolds, M.L. Wall, M. Foss-Feig,
K.R.A. Hazzard, A.M. Rey, and C.A. Regal,
Two-particle quantum interference in tunnel-coupled optical tweezers,
Science {\bf 345}, 306 (2014).
\textcolor{black}{
\bibitem{kauf18}
A.M.Kaufman, M.C. Tichy, F. Mintert, A.M. Rey and C.A.Regal,
The Hong-Ou-Mandel Effect With Atoms,
Adv. At. Mol. Opt. Phys. {\bf 67}, 377 (2018).
}
\bibitem{aspe15}
R. Lopes, A. Imanaliev, A. Aspect, M. Cheneau, D. Boiron, and C.I. Westbrook,
Atomic Hong-Ou-Mandel experiment,
Nature {\bf 520}, 66 (2015).
\bibitem{isla15}
R. Islam, R. Ma, Ph. M. Preiss, M.E. Tai, A. Lukin, M. Rispoli, and M. Greiner,
Measuring entanglement entropy in a quantum many-body system,
Nature {\bf 528}, 77 (2015).
\bibitem{bran17}
B.B. Brandt, C. Yannouleas, and U. Landman,
Two-point momentum correlations of few ultracold quasi-one-dimensional trapped fermions: 
Diffraction patterns,
Phys. Rev. A {\bf 96}, 053632 (2017).
\bibitem{bran18}
B.B. Brandt, C. Yannouleas, and U. Landman,
Interatomic interaction effects on second-order momentum correlations and Hong-Ou-Mandel interference of
double-well-trapped ultracold fermionic atoms,
Phys. Rev. A {\bf 97}, 053601 (2018).
\textcolor{black}{
\bibitem{yann19}
C. Yannouleas, B.B. Brandt, and U. Landman,
Interference, spectral momentum correlations, entanglement, and Bell inequality for a trapped 
interacting ultracold atomic dimer: Analogies with biphoton interferometry,
Phys. Rev. A {\bf 99}, 013616 (2019). 
}
\bibitem{bonn18}
M. Bonneau, W.J. Munro, K. Nemoto, and J\"{o}rg Schmiedmayer,
Characterizing twin-particle entanglement in double-well potentials,
Phys. Rev. A {\bf 98}, 033608 (2018).
\bibitem{hart06}
M.J. Hartmann, F.G.S.L. Brandao, and M.B. Plenio,
Strongly interacting polaritons in coupled arrays of cavities,
Nat. Phys. {\bf 2}, 849 (2006),
\bibitem{gree06}
A.D. Greentree, C. Tahan, J.H. Cole, and L.C.L. Hollenberg,
Quantum phase transitions of light,
Nat. Phys. {\bf 2}, 856 (2006).
\bibitem{ange07}
D.G. Angelakis, M.F. Santos, and S. Bose, 
Photonblockade-induced Mott transitions and XY spin models in coupled cavity arrays,
Phys. Rev. A {\bf 76} 031805 (2007).
\bibitem{fazi07}
D. Rossini and R. Fazio,
Mott-insulating and glassy phases of polaritons in 1D arrays of coupled cavities.
Phys. Rev. Lett. {\bf 99}, 186401 (2007).
\bibitem{brom10}
Y. Bromberg, Y. Lahini, and Y. Silberberg,
Bloch oscillations of path-entangled photons,
Phys. Rev. Lett. {\bf 105}, 263604 (2010).
\bibitem{long12}
S. Longhi and G. Della Valle,
Anyons in one-dimensional lattices: a photonic realization,
Opt. Lett. {\bf 37}, 2160 (2012).
\bibitem{lebu15}
M. Lebugle, M. Gr\"{a}fe, R. Heilman, A. Perez-Leija, S. Nolte, and A. Szameit,
Experimental observation of N00N state Bloch oscillations,
Nat. Commun. {\bf 6}, 8273 (2015).
\bibitem{ange17}
C. Noh and D.G. Angelakis, 
Quantum simulations and many-body physics with light,
Rep. Prog. Phys. {\bf 80}, 016401 (2017).
\bibitem{lein77}
J.M. Leinaas and J. Myrheim,
On the theory of identical particles,
Il Nuovo Cimento B {\bf 37}, 1 (1977).
\bibitem{wilc82}
F. Wilczek, 
Quantum mechanics of fractional-spin particles,
Phys. Rev. Lett. {\bf 49}, 957 (1982).
\bibitem{kita03}
A.Yu. Kitaev,
Fault-tolerant quantum computation by anyons,
Ann. Phys. (N.Y.) {\bf 303}, 2 (2003).
\bibitem{pach12}
J.K. Pachos,
Introduction to topological quantum computation,
(Cambridge University Press, 2012)
\bibitem{gefe12}
B. Rosenow, I. P. Levkivskyi, and B. I. Halperin, 
Current Correlations from a Mesoscopic Anyon Collider,
Phys. Rev. Lett. {\bf 116}, 156802 (2016).
\bibitem{dubc18}
T. Dub\v{c}ek, B. Klajn, R. Pezer, H. Buljan, and D. Juki\'{c},
Quasimomentum distribution and expansion of an anyonic gas,
Phys. Rev. A {\bf 97}, 011601(R) (2018).
\bibitem{umuc18}
R.O. Umucalilar, E. Macaluso, T. Comparin, and I. Carusotto,
Time-of-flight measurements as a possible method to observe anyonic statistics,
Phys. Rev. Lett. {\bf 120}, 230403 (2018).
\bibitem{keil11}
T. Keilmann, S. Lanzmich, I. McCulloch, and M. Roncaglia, 
Statistically induced phase transitions and anyons in 1D optical lattices,
Nat. Commun. {\bf 2}, 361 (2011).
\bibitem{gres15}
S. Greschner and L. Santos, 
Anyon Hubbard model in one-dimensional optical lattices,
Phys. Rev. Lett. {\bf 115}, 053002 (2015).
\bibitem{ecka16}
Ch. Str\"{a}ter, S.C.L. Srivastava, and A. Eckardt,
Floquet realization and signatures of one-dimensional anyons in an optical lattice,
Phys. Rev. Lett. {\bf 117}, 205303 (2016).
\bibitem{pels15}
G. Tang, S. Eggert, and A. Pelster,
Ground-state properties of anyons in a one-dimensional lattice,
New J. Phys. {\bf 17}, 123016 (2015).
\bibitem{fore16}
J. Arcila-Forero, R. Franco, and J. Silva-Valencia,
Critical points of the anyon-Hubbard model,
Phys. Rev. A {\bf 94}, 013611 (2016).
\bibitem{zhan17}
W. Zhang, S. Greschner, E. Fan, T.C. Scott, and Y. Zhang,
Ground-state properties of the one-dimensional unconstrained pseudo-anyon Hubbard model,
Phys. Rev. A {\bf 95}, 053614 (2017).
\textcolor{black}{
\bibitem{fesh17}
F. Lange, S. Ejima, and H. Fehske,
Anyonic Haldane Insulator in One Dimension,
Phys. Rev. Lett. {\bf 118}, 120401 (2017).
\bibitem{gors18}
F. Liu, J. R. Garrison, D.-L. Deng, Z.-X. Gong, and A. V. Gorshkov,
Asymmetric Particle Transport and Light-Cone Dynamics Induced by Anyonic Statistics,
Phys. Rev. Lett. {\bf 121}, 250404 (2018).
\bibitem{zuo18}
Z.-W. Zuo, G.-L. Li, and L. Li,
Statistically induced topological phase transitions in a one-dimensional superlattice anyon-Hubbard model,
Phys. Rev. B {\bf 97}, 115126 (2018).
\bibitem{fore18}
J. Arcila-Forero, R. Franco, and J. Silva-Valencia,
Three-body-interaction effects on the ground state of one-dimensional anyons,
Phys. Rev. A {\bf 97}, 023631 (2018).
\bibitem{mand99}
L. Mandel,
Quantum effects in one-photon and two-photon interference,
Rev. Mod. Phys. {\bf 71}, S274 (1999).
\bibitem{shihbook}
Y. H. Shih,
{\it An Introduction to Quantum Optics: Photon and Biphoton Physics\/} 
(CRC Press, Boca Raton, Florida, 2011)
\bibitem{oubook}
Z. Y. Ou,
{\it Multi-photon Quantum Interference\/}
(Springer, New York, 2007).
\bibitem{joch15}
S. Murmann, A. Bergschneider, V. M. Klinkhamer, G. Z\"{u}rn, T. Lompe, and S. Jochim,
Two fermions in a double well: Exploring a fundamental building block of the Hubbard model,
Phys. Rev. Lett. {\bf 114}, 080402 (2015).
\bibitem{prei18}
Ph. M. Preiss, J. H. Becher, R. Klemt, V. Klinkhamer, A. Bergschneider, and S. Jochim,
High-Contrast Interference of Ultracold Fermions,
Phys. Rev. Lett. {\bf 122}, 143602 (2019).
\bibitem{joch18}
A. Bergschneider, V. M. Klinkhamer, J. H. Becher, R. Klemt, L. Palm, G. Z\"{u}rn, 
S. Jochim, and Ph. M. Preiss,
Experimental characterization of two-particle entanglement through position and momentum correlations,
Nature Phys. {\bf 15}, 640 (2019), https://www.nature.com/articles/s41567-019-0508-6.
\bibitem{altm04}
E. Altman, E. Demler, and M. D. Lukin,
Probing many-body states of ultracold atoms via noise correlations,
Phys. Rev. A {\bf 70}, 013603 (2004).
\bibitem{aspe07}
T. Jeltes, J. M. McNamara, W. Hogervorst, W. Vassen, V. Krachmalnicoff, M. Schellekens, A. Perrin, 
H. Chang, D. Boiron, A. Aspect, and C. I. Westbrook,
Comparison of the Hanbury Brown-Twiss effect for bosons and fermions,
Nature {\bf 445}, 402 (2007).
\bibitem{hodg11}
S. S. Hodgman, R. G. Dall, A. G. Manning, K. G. H. Baldwin, and  A. G. Truscott,
Direct Measurement of Long-Range Third-Order Coherence in Bose-Einstein Condensates,
Science {\bf 331}, 1046 (2011).
\bibitem{hodg13}
A. G. Manning, Wu RuGway, S. S. Hodgman, R. G. Dall, K. G. H. Baldwin, and A. G. Truscott,
Third-order spatial correlations for ultracold atoms,
New J. Phys. {\bf 15}, 013042 (2013).
\bibitem{kher13}
R. G. Dall, A. G. Manning, S. S. Hodgman, Wu RuGway, K. V. Kheruntsyan, and A. G. Truscott,
Ideal $n$-body correlations with massive particles,
Nature Phys. {\bf 9}, 341 (2013).
\bibitem{kher17}
S. S. Hodgman, R. I. Khakimov, R. J. Lewis-Swan, A. G. Truscott, and K. V. Kheruntsyan,
Solving the Quantum Many-Body Problem via Correlations Measured with a Momentum Microscope,
Phys. Rev. Lett. {\bf 118}, 240402 (2017).
\bibitem{kher14}
R. J. Lewis-Swan and K. V. Kheruntsyan,
Proposal for demonstrating the Hong-Ou-Mandel effect with matter waves,
Nature Commun. {\bf 5}, 3752 (2014).
\bibitem{carn91}
O. Carnal and J. Mlynek, 
Young's double slit experiment with atoms: A simple atom interferometer, 
Phys. Rev. Lett. {\bf 66}, 2689 (1991).
\bibitem{prit09}
A. D. Cronin, J. Schmiedmayer, and D. E. Pritchard,
Optics and interferometry with atoms and molecules,
Rev. Mod. Phys. {\bf 81}, 1051 (2009).
\bibitem{bermbook}
P. R. Berman,
{\it Atom Interferomerty\/}
(Academic Press, San Diego, 1997).
\bibitem{schl15}
P. Berg, S. Abend, G. Tackmann, C. Schubert, E. Giese, W. P. Schleich, F. A. Narducci, W. Ertmer, and E. M. Rasel,
Composite-Light-Pulse Technique for High-Precision Atom Interferometry,
Phys. Rev. Lett. {\bf 114}, 063002 (2015).
}
\bibitem{bati04}
C.D. Batista and G. Ortiz,
Algebraic approach to interacting quantum systems,
Adv. Phys. {\bf 53}, 1 (2004).
\textcolor{black}{
\bibitem{note}
The spin-polarized cases ($S_z=1$, $S=1$) map to the cases of spinless particles. Namely, for bosons, there are
three basis kets $\ket{\uparrow\uparrow,0}$, $\ket{\uparrow,\uparrow}$, $\ket{0,\uparrow\uparrow}$, and the
exposition follows the case of two spinless bosons. For fermions, the fully polarized case is
trivial, because there is only one basis ket $\ket{\uparrow,\uparrow}$, leading to a single-determinantal
wave function with vanishing energy, and to a second-order momentum correlation given by Eq.\ (\ref{g4ad}). 
}
\bibitem{shih03}
Y.H. Shih,
Entangled biphoton source - property and preparation,
Rep. Prog. Phys. {\bf 66}, 1009 (2003).
\end{thebibliography}
\end{document}